\documentclass[conference]{IEEEtran}
\IEEEoverridecommandlockouts
\usepackage{cite}
\usepackage{amsmath,amssymb,amsfonts}
\usepackage{algorithmic}
\usepackage{graphicx}
\usepackage{textcomp}
\usepackage{xcolor}
\usepackage{multirow}
\usepackage{array}
\usepackage{booktabs}
\usepackage{adjustbox}
\usepackage[utf8]{inputenc}
\def\BibTeX{{\rm B\kern-.05em{\sc i\kern-.025em b}\kern-.08em
    T\kern-.1667em\lower.7ex\hbox{E}\kern-.125emX}}

\newcommand{\rom}[1]{\uppercase\expandafter{\romannumeral #1\relax}}

\begin{document}

\title{Variance of ML-based software fault predictors:\\
are we really improving fault prediction?\\
}

\author{
\IEEEauthorblockN{Xhulja Shahini}
\IEEEauthorblockA{
\textit{
Paluno, University of Duisburg-Essen} \\
Essen, Germany \\
0000-0002-5088-1614}
\and
\IEEEauthorblockN{Domenic Bubel}
\IEEEauthorblockA{\textit{
University of Duisburg-Essen} \\
Essen, Germany \\
dc.bubel@gmail.com}
\and
\IEEEauthorblockN{Andreas Metzger}
\IEEEauthorblockA{
\textit{
Paluno, University of Duisburg-Essen}\\
Essen, Germany \\
0000-0002-4808-8297}
}

\maketitle

\begin{abstract}
Software quality assurance activities become increasingly difficult as software systems become more and more complex and continuously grow in size. 
Moreover, testing becomes even more expensive when dealing with large-scale systems. 
Thus, to effectively allocate quality assurance resources, researchers have proposed fault prediction (FP) which utilizes machine learning (ML) to predict fault-prone code areas. 
However, ML algorithms typically make use of stochastic elements to increase the prediction models’ generalizability and efficiency of the training process. These stochastic elements, also known as nondeterminism-introducing (NI) factors, lead to variance in the training process and as a result, lead to variance in prediction accuracy and training time. 
This variance poses a challenge for reproducibility in research. 
More importantly, while fault prediction models may have shown good performance in the lab (e.g., often-times involving multiple runs and averaging outcomes), high variance of results can pose the risk that these models show low performance when applied in practice. 
In this work, we experimentally analyze the variance of a state-of-the-art fault prediction approach. 
Our experimental results indicate that NI factors can indeed cause considerable variance in the fault prediction models’ accuracy.
We observed a maximum variance of 10.10\% in terms of the per-class accuracy metric.
We thus, also discuss how to deal with such variance.

\end{abstract}


\section{Introduction}
Several studies have shown the increasing dependence of the industry on software systems \cite{b9, b10}.
Digitalization significantly changed the way services and products are being created and delivered, but it has also increased the crucial importance of producing high-quality software.
Software failures lead to lowered user satisfaction and high costs.

In 2022, in the US the costs of poor-quality software
were approximately \$23.35 trillion, denoting an increase of
\$2.41 trillion since 2020 \cite{b2}.
While, the usage, size, and complexity of software systems increase, so does the pressure on developers and testers to deliver high-quality software.
Thus, researchers and practitioners have a keen interest in finding efficient and effective ways to assure software quality.
Given the vast amount of data being produced by software development processes, researchers and practitioners have turned their attention to leveraging machine learning (ML) and deep learning (DL) solutions to aid software development activities. 
One of them is fault prediction as a  quality assurance task.
Fault prediction (FP) techniques leverage historical data to train ML models to predict fault-prone areas in the code \cite{b11}.
'Fault-prone' means that there is a high probability of a fault being present in that area of the code, nevertheless, there is no guarantee.  
As such fault prediction is used to effectively allocate testing resources and manual code inspection \cite{b4, b20}.
\subsection{Variance problem of ML models}
ML, and especially DL algorithms utilize stochastic elements that introduce randomness into the training process (e.g., random batch ordering, random weight
initialization) to improve generalizability and training efficiency \cite{b12}. 
However, these stochastic elements, aka nondeterminism introducing (NI) factors, introduce variance in terms of ML models' accuracy, and training runtime, among identical training runs, i.e., training runs with the exact same machine, training data, algorithm, and hyperparameters \cite{b3, b12, b13, b16}.  

Pham et al. \cite{b12} show that for DL models trained for image classification tasks, NI factors can lead to an overall accuracy variance of up to 10.80\% or even to complete misclassification of a specific data class. 
The presence of NI factors introduces several challenges, such as reproducing the results reported for a DL model.
Bouthillier et al.\cite{b3} also state that for several publications that report improvements in state-of-the-art ML approaches for general applications, the variance of the proposed approach is on the order of magnitude of the reported improvement. 
Isaksson et al.\cite{b18} point out the importance of creating stable ML/DL models by controlling their NI factors, especially in fields where the models can have a detrimental impact, like in medicine.

On one hand, reducing or eliminating the variance of ML/DL models seems to be essential for training, testing, debugging, or legal reasons\cite{b13}. 
However, on the other hand, NI factors play a crucial role in the ML/DL model's training efficiency as well as increasing its generalizability\cite{b12}. 
For e.g., developers avoid fixing seeds\footnote{Setting seeds allows developers to control non-determinism by always generating the same set of random numbers when using random number generators, see https://pytorch.org/docs/stable/notes/randomness.html} because they might hinder a model from finding the global minima and thus reduce the accuracy of the model \cite{b13}. 
Thus, we can not simply eliminate NI factors since they have their reasons for existing in the first place. 

\subsection{Consequences of Variance on Fault Prediction}
While the ML community has identified and is trying to tackle the issues caused by variance, the same can not be said for fault prediction.
We observe a lack of literature on how variance impacts fault prediction models.
Given that fault prediction models leverage ML/DL models for prediction, we can assume that these models will also exhibit variance, nevertheless, there is no evidence.
In light of this knowledge, the presence of variance of ML/DL models leads to challenges in fairly assessing and comparing the performance of different fault prediction models, not knowing if the improved accuracy is due to stochasticity or due to a better model/approach.

\subsection{Paper contributions}
This paper makes the following contributions:
\begin{itemize}
    \item \textbf{Measuring variance of fault prediction:} We measure the accuracy and runtime variance of a state-of-the-art fault prediction model DeepJIT [8]. Thereby, providing initial evidence that the variance issue is also present in FP models and not only in ML/DL benchmarks.
    \item \textbf{Analyzing the impact of different NI factors:} Using well-known faults datasets, we measure the accuracy and run-time variance that NI factors introduce for DeepJIT. Thereby, we can identify the NI factors that introduce the highest variance.
    \item \textbf{Recommendations for dealing with variance:} Even though the goal of this paper is not to provide a reproducibility framework, but rather to increase awareness, we also provide guidelines on how to deal with variance.
\end{itemize}
Our paper is structured as follows. In Chapter \ref{chap:foundations} we lay down the foundations of our work. In Chapter \ref{chap:method} we explain the methodology of our experiments, followed by the experimental results in Chapter \ref{chap:res}. In Chapter \ref{chap:diss} we discuss our findings and provide recommendations. In Chapter \ref{chap:end} we summarize our work and discuss potential future directions.

\section{Foundations} 
\label{chap:foundations}

\subsection{Fault Prediction}

Fault prediction (FP) is a static quality assurance technique. As such, FP aims to identify fault-prone code areas without actually executing the source code. 
Fault prediction, aka defect prediction/ bug prediction, leverages historical fault data and ML algorithms to train FP models.
FP takes as input different types of metrics, such as source code metrics, process metrics, developer metrics; etc. 
The output can be in the form of a) a label, that classifies if the given input is fault-prone or clean, b) a probability score that estimates the likelihood of the input being faulty, or c) the number of faults in a given input.

Fault prediction techniques can be used at different development phases.
Traditional, aka. release fault prediction, leverages fault prediction techniques to predict faults that might be present in the entire software before a new software release \cite{b4, b6}. 
Just-in-time (JIT) fault prediction, on the other side, aims at identifying faults that might be present in new change, aka. commit.
JIT fault prediction is applied every time a new commit is made to the repository \cite{b5}. 

Regarding the data used to train fault prediction models, we distinguish two main research directions, namely within-project and cross-project fault prediction.
Within-project fault prediction techniques only use data generated from the current project to train fault prediction models \cite{b19, b20}.
These techniques can only be applied in later phases of software development when enough data is available.
Cross-project fault prediction, uses data from other similar projects, to train fault prediction models that are used to predict fault-prone code areas in the current project.
The reason for using cross-project data is that during the first phases of software development, there are not enough data for training ML/DL models, or the real labels of the currently available data might not be known.
The trained fault prediction models are afterward updated with data generated from the current project, during later phases of the software development processes \cite{b8, b14, b15}.
FP is typically used before dynamic testing, to allocate testing and inspection effort on code fragments that were predicted as fault-prone.
Research suggests that FP can be used for several purposes \cite{b4, b20, b21}.
First, FP can aid project managers to assess project progress and effectively allocate resources to activities towards fault localization and fixing \cite{b4}. 
Secondly, FP helps process managers evaluate the process performance, and assess current product quality. 
Mahesh Kumar Thota et al., state that FP techniques improve quality assurance activities by providing information about the number of faults, potentially, their location and distribution in the code, or their severity \cite{b17}.
As such, FP provides useful information for project planning and risk analysis\cite{b17, b20}.
Additionally, FP aids maintenance and extension to software Just-in-time FP aims at classifying change commits as fault-prone or clean before the changes are actually committed \cite{b8}.

\textit{DeepJIT}\cite{b8} is a just-in-time FP approach. 
DeepJIT uses artifacts such as commit messages, and added and removed lines of code as input data to train an ML model, namely a convolutional neural network (CNN).
The commit messages and code changes are represented as natural language and fed into the CNN model for training.
The trained model predicts if a commit is fault-prone or not.
DeepJIT has been evaluated with two datasets, namely QT (Qt | CrossPlatform Software Development for Embedded \& Desktop, n.d.) and OPENSTACK (Open Source Cloud Computing Infrastructure, n.d.) project data. 
FP datasets typically suffer from data imbalance \cite{b5}. 
This means the percentage of data instances labeled as clean is way higher than the percentage of data instances labeled as faulty.
To deal with this imbalance, DeepJIT increases the cost of misclassifying faulty code changes. This means the FP model is penalized more when predicting an actual faulty
commit as clean and is thus more inclined to improve the model based on
this misclassification.
Hoang et al. evaluate the model using the AUC score.

\subsection{NI factors and Variance}
A software system is known to be deterministic if running it “under some fixed experimental environment, always produces identical outputs for a given input” \cite{b7}. 
Consequently, a software is nondeterministic when it produces different outputs under identical environmental conditions and input.

ML algorithms, especially DL algorithms, leverage stochastic functions and processes, i.e., non-deterministic introducing (NI) factors, to increase the efficiency and effectiveness of the training.
Some examples of NI factors are: random weight initialization, which is used to increase generalizability and avoid falling into local minima during the training of the model, and CPU/GPU parallelization which is used to increase computing efficiency \cite{b12, b13}.

Pham et al. differentiate between \textit{implementation} and \textit{algorithmic} NI factors \cite{b12}. 
The former being the NI factors pertaining to technical recourses of optimizing the efficiency training process of ML/DL models, such as parallel computing, floating-point/scheduling precision and auto-selection of primitive operations.
Algorithmic NI factors pertain to recourses of effectively training ML/DL models, such that maximum accuracy is achieved.
Some examples of algorithmic NI factors are random weight initialization, dataset shuffling/batch ordering, dropout layers, etc.
NI factors lead to accuracy variance among identically trained ML/DL models.

Variance is a measure of dispersion among the values of a variable in a data sample, denoting how close/far a specific data instance is from the expected value, i.e., from the mean and from every other instance of that data sample \cite{b22, b23}.

\section{Experimental design}
\label{chap:method}
\subsection{Methodology}
Motivated by the work of Pham et al. \cite{b12} we have used the same methodology to measure the variance in terms of accuracy and runtime of DeepJIT approach. 
We chose DeepJIT from a set of ML-based FP papers from the literature review of Wang et al. \cite{b28} because the implementation of the approach as well as the datasets are available. This is crucial for replicating the results and comparing them with the results reported in the paper. Additionally, DeepJIT paper has 100+ citations.

Pham et al. differentiate between implementation and algorithmic NI factors. 
In this work, we focus on analysing the impact of algorithmic NI factors.
To ensure complete determinism regarding implementation NI factors, would mean performing the training of an ML/DL model on a single CPU thread, which is neither realistic nor efficient \cite{b12}. 
Thus, regarding the implementation NI factors, we only analyse the impact of training fault predictors on CPU only (GPU disabled) vs. CPU+GPU (GPU enabled).
We first identify all the algorithmic NI factors present in the DeepJIT approach.
In DeepJIT we identified the following algorithmic NI factors, as can be seen in Table \ref{table:settings}: random weight initialization (W), random dropout layers (D), and random batch ordering (B). 
We experiment with all these algorithmic NI factors, and the GPU implementation NI factor, to assess the amount of variance that they cause on the accuracy and runtime of DeepJIT.
We analyse the variance of DeepJIT when none of the NI factors are turned on (Setting N), when all NI factors are turned on (Setting GA), and when each NI factor is turned on in isolation, i.e., only one NI factor is on and all the rest are off.
We turn on NI factors in isolation to quantify how much variance specific NI factors introduce into the model.
Due to a lack of resources, we do not experiment with all possible combinations of the identified NI factors.

We turn on a NI factor by setting the seed of the respective function to random. 
This means that for every run, that function will start with a random seed causing the results to be non-deterministic.
We turn off NI factors by fixing the seed of the respective function, thus forcing that function to start with a predetermined seed i.e., value, causing it to behave deterministically. 

To perform the experiments we follow these steps:
\begin{enumerate}
    \item \textit{Step 1:} Select one of the training settings (See Table \ref{table:settings}).
    \item \textit{Step 2:} Train 16 identical models and evaluate each model in terms of accuracy and runtime.
    \item \textit{Step 3:} Calculate variance among the 16 identical models, in terms of accuracy and runtime.
    \item \textit{Step 4:} Measure statistical significance using Levene's and Mann Whitney U-tests .   
\end{enumerate}

\begin{table}[!htbp]
\centering
\caption{The settings of the experiments}
\begin{tabular}{p{0.07\textwidth}|c | *3c}
\toprule
{} & {Impl. NI-factor} &  \multicolumn{3}{c}{Alg. NI factors}\\
\hline
\midrule
Setting ID & GPU & Weights init. & Dropout & Batch\\
\hline
\hline
N & Off & Off & Off & Off \\
A & Off & On & On & On \\
W & Off & On & Off & Off \\
D & Off & Off & On & Off \\
B & Off & Off & Off & On\\
\hline
GN & On & Off & Off & Off \\
GA & On & On & On & On \\
GW & On & On & Off & Off \\
GD & On & Off & On & Off \\
GB & On & Off & Off & On\\

\bottomrule
\end{tabular}
\label{table:settings}
\end{table}

Pham et al. perform 16 training runs under the same exact conditions, thus we do the same.
For each setting, we train 16 identical instances of DeepJIT and measure their accuracy and runtime.
Same as Pham et al. we measure the \textit{extreme case} and the \textit{average case} variance.
The difference between the accuracies (the highest accuracy - lowest accuracy) of these 16 trained models gives us the \textit{extreme case} of the accuracy variance, which we denote as \textit{MaxDiff} in our experiments' results.
The standard deviation of the accuracies of the 16 trained models gives us the \textit{average case} of the accuracy variance, we denote this as \textit{StdDev} in our experiments' results.
In the same way, we also calculate the runtime variance.
To confirm that these variances are statistically significant, similar to Pham et al., we perform two statistical significance tests, namely Levene's test and Mann Whitney U-test.

\subsection{Subject systems}
We train and measure the variance of DeepJIT \footnote{The data and code from DeepJIT paper can be found here https://zenodo.org/record/3965246} with the same datasets used in the DeepJIT paper.
DeepJIT has been evaluated with data from two subject systems, namely OPENSTACK 
and QT.
The QT dataset has a total of 25150
commits of which 8\% are faulty and the OPENSTACK dataset contains 12374 commits with 13\% faulty commits.
To deal with the data imbalance, the authors of DeepJIT increase the cost for misclassifying the faulty code changes \cite{b8}. 
This means the model is penalized more when predicting an actual faulty
commit as clean and is thus more inclined to improve the model based on
this misclassification.
The authors of DeepJIT have already split the data into training and testing data, and we keep the same sets for our experiments.

\subsection{Evaluation}

We evaluate our experiments at two levels.

First, we evaluate the performance of the fault prediction models, in terms of accuracy and runtime.
We use the same accuracy metrics used to evaluate the DeepJIT approach, namely the AUC score.
Additionally, we also compute for each trained model the confusion matrix metrics, to help us better understand and interpret the results, and the per-class accuracy (i.e., the number of times the model correctly classifies the specific class), following Pham et al.
We also measure the total training time, to be able to answer our second research question.
Comparing the differences in AUC scores and per-class accuracies of the 16 identical training runs, i.e., fault prediction models trained under the exact same setting gives us the variance of that specific setting.

Second, for each setting, we compute the statistical significance of the measured variance via the Levene's and Mann Whitney U-tests, to confirm if the observed variance is statistically significant. 
We compute the statistical significance of the variance caused by each algorithmic NI factors in isolation, as well as all NI factors altogether: by comparing the results of each setting (W, D, B, and A) with (setting N) for experiments with GPU off.
Similarly, we compare each setting, namely (GW, GD, GB, and GA) with (Setting GN) for experiments with GPU on.
We compute the statistical significance of the variance caused by the implementation-level NI factors, namely GPU parallelization, by comparing the results of each setting with GPU on, vs the corresponding setting with GPU off. 
E.g., the variance caused by the random weight initialization (Setting W) when training 16 identical models with GPU support off, vs (Setting GW) when training 16 identical models with GPU support on.

\section{Experimental results}
\label{chap:res}

We ran all our CPU-only (GPU-disabled) experiments on the MagnitUDE supercomputer provided by the University of Duisburg-Essen.
The MagnitUDE supercomputer operates on Linux and offers multiple computing nodes with 24 cores each.
Unfortunately, Magnitude does not offer GPU support, thus, for the CPU+GPU (GPU-enabled) experiments, we used a AMD Ryzen 5 1600 six-core (CPU), GeForce GTX 1060 6GB (GPU), 16 GB of DDR4 RAM computer.
During the training of the models, the computer has not been used for any other processes to ensure no interference or runtime delays. 
The library versions we used are: PyTorch 1.12.1 (CUDA 11.3), tqdm 4.64.0, numpy 1.23.1, scikit-learn 1.1.2.
In total, we conducted 162 (2 were test runs) experiments on GPU and 160 experiments on CPU, with a total of 421.33 hours of CPU+GPU training on a private computer and 581.63 hours of CPU-only training on MagnitUDE. The code and experiments can be found in our Github repository\footnote{ Detailed information regarding used code and exact results are available in the Git repository: https://github.com/Domenic-7/variance\_analysis\_deepjit }.

\subsection{How much accuracy variance do NI factors
introduce for the given fault prediction approach (DeepJIT)}
\subsubsection{OPENSTACK}
Table \ref{table:Open-GPU} and \ref{table:Open-CPU} display the results of our experiments when training DeepJIT with OPENSTACK dataset, with GPU-enabled and GPU-disabled respectively.
We display both MaxDiff and StdDev for the AUC score, per-class accuracy for faulty commits, and per-class accuracy for clean commits.
The numbers in bold format, represent the highest variance within the respective column.

We can observe from Table \ref{table:Open-GPU}, that there is accuracy variance for each of the settings, with a lowest measured variance of 0.12\% and a maximum of 10.10\%. 
Looking at the different settings, we can easily identify that random weight initialization (Setting GW) is the NI factor introducing the highest variance for all the metrics.
The same observation was also made by Pham et al.
Random weight initialization has such a big variance impact because it affects the starting point of the DL model.
Meanwhile, dropout-layer, and random batch ordering, as well as implementation level NI factors, influence how the model adjusts during the training.
Note that the effect of all NI factors on variance is not equal to the sum of the effect of every NI factor in isolation, because these effects are not independent of each-other \cite{b3}.
Also note that the GPU is enabled for these experiments, which means that the measured effect of (Setting GW) might actually be a combination of the variance caused by GPU parallelization and random weight initialization.
We can only observe the isolated effect of random weight initialization when the GPU is disabled, see Table \ref{table:Open-CPU}.
Meanwhile, the (Setting GN), corresponding to the effect of GPU parallelization, introduces the least accuracy variance.
This observation matches the results of \cite{b7, b12},  where the implementation-level NI factors led to the lowest variance out of all NI factors.
Looking at the maximum differences and standard deviations in AUC score caused by dropout (Setting GD) and random batch ordering (Setting GB), it seems that they both introduce similar variance. 
However, when looking at per-class accuracy we can notice that the random batch ordering led to a standard deviation of over 2x compared to dropout-layer for faulty commits and over 1.5x for clean commits. 
The calculated AUC scores did not represent this difference, and it varies much less than the per-class accuracies. 
To investigate why the difference in magnitude of variance between the AUC score and the per-class accuracies, we also computed the confusion matrix.
We understood that the reason why we observe this difference is that the per-class accuracy metric is measured using a consistent threshold of 0.5, while the AUC score is calculated across all possible thresholds. 
This is because the AUC score is calculated from the ROC curve which plots the true positive rate and false negative rate for every possible classification threshold, thus it is independent of a threshold and gives a general overview of the model's accuracy.
Jiang et al. also reported that the AUC score is more robust to variance \cite{b24}.
This brings us to the conclusion that accuracy metrics should be picked carefully, in order to reflect how much the model has learned during training.

Levene's statical tests confirmed the significance of the observed variances. 
The tests are calculated by comparing the AUC score of each Setting (GA, GW, GD, GB) with the AUC score of (Setting GN).
The tests proved that all the different settings caused statistically significant variance.
Mann Whitney U-tests also confirmed that the AUC score and per-class accuracy variance introduced by every NI-factor is statistically significant,
Except for the AUC score of the (Setting GW), which did not reject the null hypothesis.\\

\noindent\fbox{%
\parbox{0.48\textwidth}{%
\textit{Conclusion: When training DeepJIT on GPU, using OPENSTACK dataset, we observed a maximum accuracy variance of 10.10\%. The statistical tests confirmed the significance of the variance caused by all NI factors combined, as well as each NI factor in isolation.}
}%
}\\

\begin{table}[!htbp]
\centering
\caption{ DeepJIT Variance - OPENSTACK dataset - GPU enabled}
\begin{adjustbox}{width=0.47\textwidth}
\begin{tabular}{ m{2em}  *6c}
\toprule
{}&  \multicolumn{2}{c}{AUC score (\%)} & \multicolumn{2}{c}{per-class: faulty(\%)} & \multicolumn{2}{c}{per-class: clean(\%) }\\
\midrule
Setting  & MaxDiff  & StdDev & MaxDiff  & StdDev & MaxDiff  & StdDev\\
\hline
\hline
GN & 0.12 & 0.04 & 0.61 & 0.27 & 0.34 & 0.10 \\
GA & 1.45 & 0.38 & 7.36 & 2.20 & 8.90 & 2.50 \\
GW & \textbf{1.48} & \textbf{0.45} & \textbf{8.59} & \textbf{2.55} & \textbf{10.10} & \textbf{2.69} \\
GD & 0.53 & 0.13 & 1.84 & 0.59 & 2.74 & 0.72 \\
GB & 0.54 & 0.16 & 4.29 & 1.36 & 4.20 & 1.16 \\
\bottomrule
\end{tabular}
\end{adjustbox}
\label{table:Open-GPU}
\end{table} 

\begin{table}[!htbp]
\centering
\caption{DeepJIT Variance - OPENSTACK dataset - GPU disabled}
\begin{adjustbox}{width=0.48\textwidth}
\begin{tabular}{ m{2em}  *6c}
\toprule
{}&  \multicolumn{2}{c}{AUC score (\%)} & \multicolumn{2}{c}{per-class: faulty(\%)} & \multicolumn{2}{c}{per-class: clean(\%) }\\
\midrule
Setting  & MaxDiff  & StdDev & MaxDiff  & StdDev & MaxDiff  & StdDev\\
\hline
\hline
N & 0 & 0 & 0 & 0 & 0 & 0 \\
A & \textbf{0.97} & \textbf{0.33} & 3.07 & 0.96 & \textbf{5.91} & \textbf{1.94} \\
W & 0.87 & 0.24 & \textbf{6.13} & \textbf{1.58} & 5.14 & 1.43 \\
D & 0.28 & 0.10 & 2.45 & 0.74 & 2.91 & 0.77 \\
B & 0.92 & 0.28 & 2.45 & 0.80 & 1.88 & 0.60 \\
\bottomrule
\end{tabular}
\end{adjustbox}
\label{table:Open-CPU}
\end{table}

In Table \ref{table:Open-CPU} we can observe that the total effect of all algorithmic NI factors (Setting A) is usually bigger than the effect of random weight initialization alone (Setting W).
In these experiments we can notice that the effect of (Setting W) is still the highest compared to the other algorithmic NI factors in isolation, nevertheless, compared to the experiments with GPU-enabled, the magnitude of this effect is lower.
This can be explained by the fact that, by turning off the GPU support, we eliminate the combined effect of random weight initialization and GPU parallelization and we only see the effect of random weight initialization.
As we can see, the (Setting N) has consistently zero effect, which is expected, given that we have turned off the GPU support and we ran these experiments only on CPU. 
To evaluate the statistical significance of the variance introduced by implementation-level NI factors, we computed Levene’s test between the AUC scores of the GPU-enabled and GPU-disabled experiments.
For (W vs GW) and (B vs GB) settings  Levene’s test rejected the null hypothesis and thus confirmed that implementation-level NI factors introduced significantly different variances. 
However, for (A vs GA) and (D vs GD) settings the statistical significance test accepts the null hypothesis, meaning, implementation-level NI factors introduced no statistically significant variance across GPU and CPU experiments. 
However, regarding setting (A vs GA), the test rejects the null hypothesis comparing the variance of per-class accuracy of faulty commits but accepts it for the per-class accuracy of clean commits.\\

\noindent\fbox{%
\parbox{0.48\textwidth}{%
\textit{Conclusion: When comparing training DeepJIT on CPU vs on GPU, using OPENSTACK dataset, we obtain inconsistent results regarding the statistical significance of the variance.}
}%
}

\begin{table}[!htbp]
\centering
\caption{DeepJIT Variance - QT dataset - GPU enabled}
\begin{adjustbox}{width=0.48\textwidth}
\begin{tabular}{ m{2em}  *6c}
\toprule
{}&  \multicolumn{2}{c}{AUC score (\%)} & \multicolumn{2}{c}{per-class: faulty(\%)} & \multicolumn{2}{c}{per-class: clean(\%) }\\
\midrule
Setting  & MaxDiff  & StdDev & MaxDiff  & StdDev & MaxDiff  & StdDev\\
\hline
\hline
GN & 0.06 & 0.02 & 1.09 & 0.40 & 0.50 & 0.14 \\
GA & \textbf{1.38} & 0.36 & \textbf{8.20} & \textbf{2.12} & \textbf{6.78} & \textbf{2.04} \\
GW & 1.34 & \textbf{0.41} & 6.01 & 1.75 & 4.86 & 1.51 \\
GD & 0.26 & 0.07 & 4.37 & 1.16 & 2.64 & 0.75 \\
GB & 0.50 & 0.12 & 5.46 & 1.76 & 3.31 & 0.90 \\
\bottomrule
\end{tabular}
\end{adjustbox}
\label{table:QT-GPU}
\end{table}

\begin{table}[!htbp]
\centering
\caption{DeepJIT Variance - QT dataset - GPU disabled}
\begin{adjustbox}{width=0.48\textwidth}
\begin{tabular}{ m{2em}  *6c}
\toprule
{}&  \multicolumn{2}{c}{AUC score (\%)} & \multicolumn{2}{c}{per-class: faulty(\%)} & \multicolumn{2}{c}{per-class: clean(\%) }\\
\midrule
Setting  & MaxDiff  & StdDev & MaxDiff  & StdDev & MaxDiff  & StdDev\\
\hline
\hline
N & 0 & 0 & 0 & 0 & 0 & 0 \\
A & 0.99 & \textbf{0.29} & 6.01 & 2.12 & \textbf{6.32} & \textbf{1.89} \\
W & \textbf{1.02} & \textbf{0.29} & \textbf{10.38} & \textbf{2.72}  & 5.82 & 1.88 \\
D & 0.29 & 0.09 & 4.37 & 1.13 & 3.52 & 0.79 \\
B & 0.29 & 0.08 & 4.92 & 1.32 & 3.43 & 0.95 \\
\bottomrule
\end{tabular}
\end{adjustbox}
\label{table:QT-CPU}
\end{table}
\subsubsection{QT}
Table \ref{table:QT-GPU} and \ref{table:QT-CPU} display the results of our experiments when training DeepJIT with QT dataset, with GPU-enabled and GPU-disabled respectively.
Same as for the experiments held using the OPENSTACK dataset, the settings (A) and (W) for GPU disabled, and settings (GA) and (GW) for GPU enabled, are the ones introducing the highest accuracy variance.
The maximum variance that we observe is 8.20\% for the GPU-enabled experiments and 10.38\% for the GPU-disabled experiments.
Once again, the per-class accuracy metrics display higher variance than the AUC score. 
Concerning Levene’s tests for the QT dataset on GPU, 
the tests confirmed that all NI factors together (Setting GA) caused statistically significant variance.
Similar to the OPENSTACK experiments, Levene’s tests also confirmed the significance of the variance introduced by every other setting.
Mann Whitney U-test also resulted in statistically significant variance caused by each setting.\\

\noindent\fbox{%
\parbox{0.48\textwidth}{%
\textit{Conclusion: When training DeepJIT on GPU, using QT dataset,  we observed a maximum accuracy variance of 8.20\%. The statistical tests confirmed the significance of the variance caused by all NI factors combined, as well as each NI factor in isolation.}
}%
}\\

When computing the statistical significance of implementation-level factors between the GPU-enabled and GPU-disabled experiments, Levene's tests accepted the null hypothesis for each setting.
This means that the variance, in terms of AUC score, introduced by implementation-level NI factors is not statistically significant. 
We also conducted Levene’s test on the variance in terms of per-class accuracy metric, because the AUC score showed less variance overall which might impact the statistical tests. However, the results are almost the same, with the tests accepting the null hypothesis in all cases but setting (W vs GW) for the per-class accuracy of faulty commits. 
Meaning, in the setting (W vs GW), implementation-level NI factors led to significantly different variances in the accuracy of faulty commits.\\

\noindent\fbox{%
\parbox{0.48\textwidth}{%
\textit{Conclusion: When comparing training DeepJIT on CPU vs on GPU, using QT dataset, we only observe statistically significant variance for random weight initialization in terms of per-class accuracy of faulty commits.}
}%
}\\

\subsection{How much runtime variance do NI factors
introduce for the given fault prediction approach (DeepJIT)}

Table \ref{table:runtime-variance} displays the mean and the standard deviation (StdDev) of the runtime of training the models with the OPENSTACK and QT datasets, with GPU-enabled and GPU-disabled respectively, for each experimental setting.

\begin{table}[!htbp]
\centering
\caption{Training time in terms of minutes}
\begin{tabular}{*7c}
\toprule
& & \multicolumn{2}{c}{OPENSTACK} & & \multicolumn{2}{c}{QT }\\
\midrule
Setting  & & Mean   & StdDev &   & Mean   & StdDev\\
\hline
\hline

N & & 143. 95 & 4.43 & &292.44  & 0.82\\
A & & 143.33 & 3.29 & &292.43 & 0.81\\
W & & 143.56  & 3.04 & &293.11  & 1.66\\
D & & 143.05  & 0.85 & &292.92  & 1.58\\
B & & 143.51  & 3.69 & &292.83  & 0.90\\
\hline
GN & & 105.30 & 0.31 & &211.01 & 0.15 \\
GA & & 105.21 & 0.28 & &210.48 & 0.69 \\
GW & & 105.28 & 0.25 & &210.85 & 0.32 \\
GD & & 105.30 & 0.28 & &210.66 & 0.54 \\
GB & & 105.23 & 0.26 & &210.65 & 0.52 \\
\bottomrule
\end{tabular}
\label{table:runtime-variance}
\end{table}
\subsubsection{OPENSTACK}
Regarding the experiments on OPENSTACK dataset, the mean training time was constantly  $\sim$105 minutes with GPU-enabled, and $\sim$143 minutes with GPU-disabled. 
Levene’s test could not reject the null hypothesis.
Thus, we can say that the NI factors did not introduce a statistically significant variance in terms of training time.
\subsubsection{QT}
Regarding the experiments on the QT dataset, the mean training time was constantly $\sim$210 minutes with GPU-enabled, and $\sim$292 minutes with GPU-disabled. 
The difference between the lowest and highest mean training time is only about half a minute. 
In this case, Levene’s test confirmed the statistical significance of the variance introduced by NI factors. \\

\noindent\fbox{%
\parbox{0.48\textwidth}{%
\textit{Conclusion: regarding runtime variance, we obtained contradictory results between the two datasets.}
}%
}\\

\subsection{Validity risks}

Concerning internal validity, one of the biggest risks is not having identified all NI factors in the code of DeepJIT. 
To make sure we have, we turned off all the identified NI factors and run the code 16 times to measure variance. This experiment resulted in zero variance, which means there is no unidentified NI factors. See (Setting N) in Table \ref{table:Open-GPU} and \ref{table:QT-GPU}.
We also carefully reviewed the implementation, to make sure there are no faults in it. 
We can not make sure there are no faults in the implementation of the third-party libraries used in DeepJIT, nevertheless, they should not affect the variance analysis we performed.

Concerning external validity, we can not say that our analysis would result in the same outcomes if performed on other fault prediction approaches. More experiments with other FP approaches and datasets should be performed in order to be able to generalize our results.

\section{Discussion and Recommendations}
\label{chap:diss}
\subsection{Discussion of insights}
As reported above, NI factors can have a significant impact on the accuracy of fault prediction models. 
As also stated by \cite{b3}, we noticed that the accuracy variance introduced by NI factors can be even larger than the accuracy improvements reported by the newly proposed approaches, in our case DeepJIT. 
Random weight initialization led to the highest variance across both GPU and CPU experiments on both datasets. 
This was followed by random batch ordering, dropout-layer, and lastly by the implementation-level NI factors. 

\begin{figure}[h]
\includegraphics[width=0.4\textwidth]{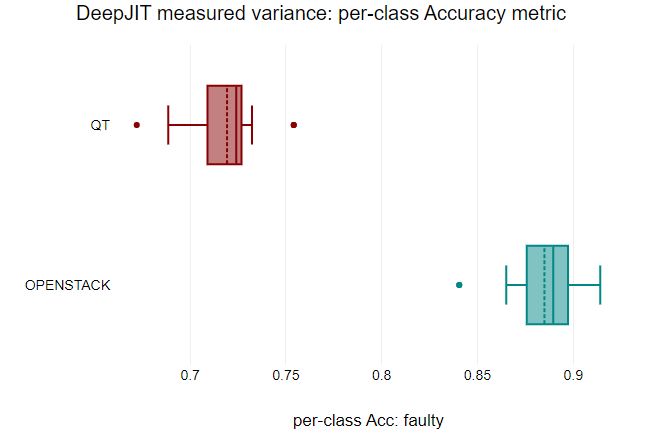}
\includegraphics[width=0.4\textwidth]{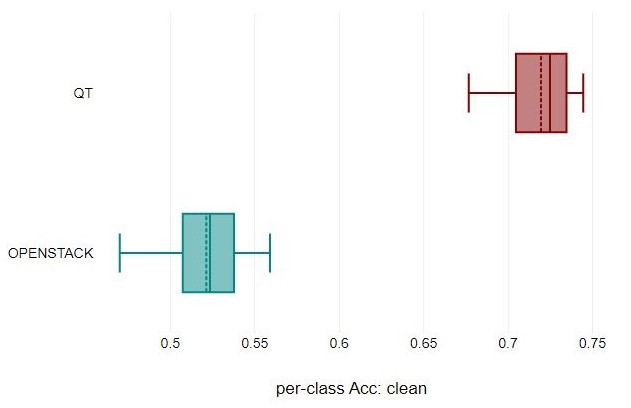}
\caption{Boxplots of the measured variance in terms of \textit{per-class accuracy: clean} and \textit{per-class accuracy: faulty} commits.}
\label{fig: faulty}
\end{figure}

Hoang et al.\cite{b8} report an AUC score of 75.1\% for the
OPENSTACK dataset and 76.8\% for the QT dataset. 
The authors compare DeepJIT accuracy scores to other state-of-the-art FP approaches such as JIT with an AUC score of 69.1\% and DBNJIT with an AUC score of 69.4\%.
Hoang et al. report that DeepJIT improves the best-performing SOTA approach by 5.7\% on AUC score for the OPENSTACK dataset and 6.3\% for the QT dataset. 
A later study by Zeng et al. \cite{b26} found contradictory results in which DeepJIT could not consistently outperform other approaches. 
Furthermore, we found that NI factors can introduce a variance in terms of AUC score of up to 1.45\% for the OPENSTACK dataset and 1.38\% for the QT dataset. 
Moreover, the introduced variance can be larger than 10\% for the per-class accuracies, see Fig \ref{fig: faulty}.
This raises the question: "Is the improved accuracy due to a better approach, or due to variance ?"

\subsection{Recommendations for dealing with variance in fault prediction}
Our findings suggest that it is difficult to compare different fault prediction approaches in a fairly, because we do not know for sure if a newly proposed approach is better, or the improvement of accuracy is only a chance result due to the NI factors.
Yet again, a good percentage of researchers are not even aware of it \cite{b3}. 
Pham et al. performed a survey in which they conclude that only around 17\% out of 901 surveyed researchers and practitioners are aware of the impact of the NI factors on the ML/DL model's accuracy.
Regarding research publications published in top-tier artificial intelligence (AI) and software engineering (SE) conferences, the authors identified only about 19.5\% publications that measure the variance of their proposed ML/DL approaches.

While turning off the NI factors to avoid variance might not be a realistic choice, since NI factors play a very important role in the efficiency and effectivity of training ML/DL models, measuring the variance at least allows reporting results in a more transparent way. 
For e.g., instead of reporting only the highest accuracy results of the ML/DL model, the authors might also report the results of running the model several times with the same environment and hyper-parameters, and report all the achieved accuracies among these runs. 
Alternatively, one might report the performance of the proposed approach via the mean and the standard deviation of the accuracy. 
As we saw from our experiments, it is crucial to properly pick the accuracy metrics, or even better, to use several different metrics that illustrate the performance of the model.
Of course, performing all these experiments comes with increased costs, but they lay down a path for better comparison of results.

Another approach for properly reporting the results of fault prediction approaches in a comparable way is to build a common benchmark.
Such a benchmark could serve for training and evaluating fault prediction models, under the same environmental conditions, datasets,  accuracy metrics, and hyper-parameter tunning techniques, for e.g., grid-search, etc. 
The benchmark should also compute the variance of the trained model.
Bouthillier et al. \cite{b3} suggest the same approach for reporting ML/DL models' results.
In the same work, they suggest randomizing as many NI factors as possible, including pre-processing and training NI factors, to minimize the standard error of the model's accuracy.

Pham et al. built a tool named Deviate \cite{b25} as a follow-up work for the paper "Problems and opportunities in training deep learning software systems: An analysis of variance".
Deviate only measures the variance of a model among 8 identical training runs.
Deviate needs human support to identify or confirm the NI factors present in a given approach. 
Although the goal of Deviate is only to measure the variance of a given ML/DL model, rather than constructing a benchmark for fair comparison of ML/DL models, such a tool could be used as a basis for building a comparison benchmark for fault predictors. 

\section{Related Work}

Assessing and controlling variance in fault prediction models has not received much interest in the literature.

Jiang et al. \cite{b24} appear to be the only authors that measure the impact a fault prediction model’s accuracy variance of using different a) classifiers, i.e., ML algorithms, b) accuracy metrics, c) dataset sizes, d) training set sizes, and e) number of 10-fold cross-validation repetitions. 
In this study, the authors conclude different ML algorithms display different degrees of variance.
Another important observation is that smaller datasets introduce more variance. 
When dealing with small datasets, also the number of 10-fold cross-validation repetitions matters, thus, one should aim to perform as many repetitions as possible to minimize variance.
Regarding the training set size, one can minimize variance by balancing the sizes of training and test sets. 
This study was published in 2009, and it only analyses traditional ML models.
We, explicitly consider a much more recent DL-based fault prediction approach, DeepJIT.
Nevertheless, the focus of our paper is different.
We do not focus on measuring how much variance do different ML models introduce.
Our work consists of measuring the accuracy as well as the runtime variance of a specific state-of-the-art fault prediction approach in terms of algorithmic and implementation NI factors.
Additionally, we also decompose the effect of different NI factors in isolation on the accuracy and runtime variance.

\section{Conclusion and Outlook}
\label{chap:end}

In this work, we conducted several experiments with the state-of-the-art fault prediction approach DeepJIT, to show the impact of NI factors on the variance, in terms of accuracy and runtime, of identically trained models. 
Our experiments showed that NI factors caused a maximum difference of up to 1.48\% in the AUC score, 8.59\% in the per-class accuracy for faulty change commits, and 10.10\% in the accuracy for clean change commits. 
Regarding runtime variance, we did not obtain consistent results among experiments, thus, we cannot say for sure if NI factors affect DeepJIT model's training runtime.
We also investigated the impact of each NI factor in isolation to the accuracy variance of DeepJIT. 
Each algorithmic NI factor in isolation, namely the random weight initialization, the dropout layer, the random batch order, and the implementation-level NI factors (GPU parallelization) introduces significant variance in the training process of DeepJIT. 
Our experimental results raise questions about the validity of the evaluations of fault prediction approaches and if the reported improvements are due to better approaches or due to the NI factors-related variance. 
Thus, we strongly suggest, researchers should consider the variance when reporting the performance of the proposed ML/DL-based FP techniques.

Our future work consists of:
\begin{enumerate}
    \item Assessing the variance of other fault prediction approaches, including approaches that use DL algorithms, as well as approaches that use simple ML algorithms such as decision trees and logistic regression.
    \item Analysing the variance introduced by other NI factors, such as those that are part of pre-processing the ML/DL model: over-sampling methods, dataset split, etc.
    \item Analysing variance when training fault predictors with other datasets, of different sizes and percentages of faults, to see if the size of a dataset affects variance.
    \item Additionally we can also try to minimize variance, e.g., by using bagging algorithms, as suggested by Zhou \cite{b27}.
\end{enumerate}

\vspace{12pt}

\begin{thebibliography}{00}
\bibitem{b1} G. Eason, B. Noble, and I. N. Sneddon, ``On certain integrals of Lipschitz-Hankel type involving products of Bessel functions,'' Phil. Trans. Roy. Soc. London, vol. A247, pp. 529--551, April 1955.

\bibitem{b2} H. Krasner, ''Cost of poor software quality in the U.S.: A 2022 REPORT'', Consortium for Information and Software Quality (CISQ)

\bibitem{b3} X. Bouthillier, P. Delaunay, M. Bronzi, A. Trofimov, B. Nichyporuk, J. Szeto, N. M. Sepahvand, E. Raff, K. Madan, V. Voleti, S. E. Kahou, V. Michalski, T. Arbel, C. Pal, G. Varoquaux, P. Vincent, ``Accounting for Variance in Machine Learning Benchmarks.'' Proceedings of Machine Learning and Systems 3, 2021. 



\bibitem{b4} S. Pandey, R. B. Mishra, and 
 A. K. Tripathi,'' ML Based Methods for Software Fault Prediction: A Survey,''  Expert Sys. with Appl., 2021.

 \bibitem{b5} Y. Zhao, K. Damevski, and Hui Chen, ''A Systematic Survey of Just-in-Time Software Defect Prediction, '' ACM Comput. Surv. 55, 2023. 

 \bibitem{b6} N. C. Shrikanth, S. Majumder, and T. Menzies, ''Early Life Cycle Software Defect Prediction. Why? How? '', In 43rd IEEE/ACM
International Conference on Software Engineering, ICSE 2021, 22-30 May 2021.

\bibitem{b7} P. Nagarajan, G. Warnell, and P. Stone, '' Deterministic Implementations for Reproducibility in Deep Reinforcement Learning, '' 2019.

\bibitem{b8} T. Hoang, H. Dam Khanh, Y. Kamei, D. Lo, and N. Ubayashi, ''
DeepJIT: An End-to-End Deep Learning Framework for Just-in-Time Defect Prediction, '' IEEE/ACM 16th International Conference on Mining Software Repositories (MSR), 34–45, 2019.

\bibitem{b9} A. Arora, L. G. Branstetter, M. Drev, '' Going Soft: How the Rise of Software-Based Innovation Led to the Decline of Japan's IT Industry and the Resurgence of Silicon Valley'', The Review of Economics and Statistics, Vol95 (3), pp 757–775, 2013.

\bibitem{b10} B. Beckert, K. Schleife, D. Dupuis, '' The economic and social impact of software \& services on competitiveness and innovation : final report, '' European Commission, Directorate-General for Communications Networks, Content and Technology,, Publications Office, 2017

\bibitem{b11} Meiliana, S. Karim, H. L. H. S. Warnars, F. L. Gaol, E. Abdurachman and B. Soewito '' Software metrics for fault prediction using machine learning approaches: A literature review with PROMISE
repository dataset, '' IEEE International Conference on
Cybernetics and Computational Intelligence (CyberneticsCom), 19–
23, 2017.

\bibitem{b12}  H. V. Pham, S. Qian, J. Wang, T. Lutellier, J. Rosenthal, L. Tan,Y. Yu, and N. Nagappan,'' Problems and opportunities in training deep learning software systems: An analysis of variance '',  Proceedings of the 35th IEEE/ACM ASE, 771–783, 2020.

\bibitem{b13} B. Chen, M. Wen, Y. Shi, D. Lin, G. K. Rajbahadur, and Z. M. Jiang, '' Towards Training Reproducible Deep Learning Models '', ICSE, 2022.

\bibitem{b14} Y. Jiang, J. Lin, B. Cukic, and T. Menzies, '' Variance Analysis in Software Fault Prediction Models '', 20th International
Symposium on Software Reliability Engineering, 99–108, 2009.

\bibitem{b15} S. Kim, E. J. Whitehead, and Y. Zhang, '' Classifying Software Changes: Clean or Buggy? '',  IEEE Transactions on Software
Engineering, Vol 34(2), pp181–196, 2008.

\bibitem{b16} C. Colas, O. Sigaud, and P. Y. Oudeyer,'' How Many Random Seeds? ``, Statistical Power Analysis in Deep RL Experiments, 2018.

\bibitem{b17} M. K. Thota, F. H. Shajin, and P. Rajesh, '' Survey on software defect prediction techniques'', International Journal of Applied Science and Engineering, Vol17(4), pp331–344, 2020.

\bibitem{b18} A. Isaksson, M. Wallman, H. Göransson, and M. G. Gustafsson, ''Cross-validation and bootstrapping are unreliable in small sample classification'', Pattern Recognition Letters, Vol29(14), pp1960–1965, 2008.

\bibitem{b19} K. Yogita and K. S. Sandeep, '' Cross project defect prediction: a comprehensive survey with its SWOT analysis'', Innov. Syst. Softw. Eng. 18, Vol 2, pp 263–281, 2022.

\bibitem{b20} J. Pachouly, S. Ahirrao, K. Kotecha, G. Selvachandran, and A. Abraham, ''A systematic literature review on software defect prediction using artificial intelligence: Datasets, Data Validation Methods, Approaches, and Tools '', Eng. Appl. Artif. Intell. 111, 2022.

\bibitem{b21} B. Clark, and D. Zubrow, D, ''How Good Is the Software: A Review of Defect Prediction Techniques '', Data Science: From Research to Application, 2001.

\bibitem{b22} Y. Zhang, H. Wu and L. Cheng, "Some new deformation formulas about variance and covariance," Proceedings of International Conference on Modelling, Identification and Control, pp. 987-992, 2012.

\bibitem{b23} D. V. Huntsberger and P. Billingsley, ''Elements of Statistical Inference'', Allyn and Bacon Inc., 1981.

\bibitem{b24}Y. Jiang, J. Lin, B. Cukic, and T. Menzies, ''Variance analysis in software fault prediction models'', In Proceedings of the 20th IEEE international conference on software reliability engineering (ISSRE'09), 2009.

\bibitem{b25} H. V. Pham, M. Kim, L. Tan, Y. Yu and N. Nagappan, "DEVIATE: A Deep Learning Variance Testing Framework," 36th IEEE/ACM Int. Conf. on Automated Software Engineering (ASE), pp. 1286-1290, 2021.

\bibitem{b26} Z. Zeng, Y. Zhang, H. Zhang, L.Zhang, "Deep just-in-time
defect prediction: How far are we?", Proceedings of the 30th ACM SIGSOFT International Symposium on Software Testing and Analysis, 427–438,

 \bibitem{b27} Zhou, Z.-H,."Ensemble Methods: Foundations and Algorithms (1st ed.)", Chapman and Hall/CRC, 2012.

 \bibitem{b28} S. Wang L. Huang, A. Gao, J. Ge, T. Zhang, H. Feng, I. Satyarth, M. Li, H. Zhang, and V. Ng. " Machine/Deep Learning for Software Engineering: A Systematic Literature Review", IEEE Transactions on Software Engineering early access, 2022,


\end{thebibliography}
\end{document}